\documentclass[fleqn,10pt]{wlscirep}
\usepackage[utf8]{inputenc}
\usepackage[T1]{fontenc}
\usepackage{amsmath, amsthm, amssymb, amsfonts}

\begin{document}

\title{Slow inter-minima relaxation and its consequence for BEC of magnons} 
\author[1*]{Gang Li}
\author[2]{Haichen Jia}
\author[1,3]{Valery Pokrovsky}
\affil[1]{Department of Physics \& Astronomy, Texas A\&M University, College Station, Texas 77843-4242, USA}
\affil[2]{Department of Physics, The University of Hong Kong, Hong Kong, China}
\affil[3]{Landau Institute for Theoretical Physics, Chernogolovka, Moscow District 142432, Russia}

\affil[*]{dgzy03@gmail.com}

%\keywords{Keyword1, Keyword2, Keyword3}

\begin{abstract}
Two recent articles of the Münster University
experimental team led by S.O. Demokritov \cite{Borisenko 2020-1, Borisenko 2020-2}
displayed several important facts related to the Bose-Einstein condensation of magnons (BECM) under permanent pumping first discovered  in 2006 \cite{Demokritov 2006}.  They contradicted existing theories of this phenomenon, which predict the attractive interaction between magnons \cite{Tupitsyn 2008, Rezende 2009, Li 2013} and strong spontaneous violation of the reflection symmetry \cite{Li 2013}. In this article, we show that these theories implicitly assumed all relaxation processes to be fast compared 
with the lifetime of the magnons, whereas one of them -- relaxation between two minima of energy -- is slow. 
We classify processes responsible for the inter-minima relaxation and present their analytic theory. We analyze
how the slow inter-minima relaxation modifies the anticipated properties of a ferromagnet with the magnon condensate. \end{abstract}

\maketitle
%\flushbottom

%\thispagestyle{empty}

\subsection*{Introduction} 
The Bose-Einstein condensation of magnons was found in a 5 $\mu m$ thick film of Yttrium Iron Garnet (YIG) \cite{Demokritov 2006}. The frequency $\omega$ of a magnon in such a film depends on the in-plane momentum $\mathbf{k}$ and the wave-vector $k_{\perp}$ of a standing way $\cos (k_{\perp} x)$ describing the distribution of magnetization within the film. Further we accept the direction of spontaneous magnetization for $z$-axis, the direction of the second component of the in-plane wave vector for $y$-axis as shown in Fig. 1a.  

Let $-J$ be the exchange interaction energy of the nearest spins, $a$ be the lattice constant and $Z$ be the coordination number. The characteristic dipolar interaction is $E_{dip}=\mu_B^2/a^3$, where $\mu_B$ is the Bohr's magneton. The so-called dipolar length $\ell = \sqrt{JZ/E_{dip}}a$ is an important length scale for magnets: at distances smaller than $\ell $ the dipolar interaction dominates, at larger the exchange interaction prevails. In YIG $\ell\approx 40  nm$\cite{Li 2018}. In thick films $d\gg\ell$, the competition of the exchange and dipolar interactions leads to the magnons spectrum with two symmetric minima at momenta $\mathbf{k} = \pm \mathbf{Q}$ parallel and antiparallel to the $z$-axis and $k_x\approx \pi/d$ \cite{Sonin 2017} and maximum of energy at $\mathbf{k}=0$. The magnitude of $\mathbf{Q}$ is $Q=\frac{(2\pi^3)^{1/4}}{\sqrt{d\ell}}$\cite{Sonin 2017}.  Magnetic field creates a Zeeman gap in the spectrum so that the frequency in the two minima of energy is equal to $\omega (\pm \mathbf{Q})=\gamma H$, where $\gamma = e/mc$ is the doubled gyromagnetic factor. The maximum frequency $\omega(0)=\gamma\sqrt{H(H+4\pi M)}$ corresponds to the ferromagnetic resonance. In 2012 the M\"unster group confirmed the existence of the two condensates finding the interference stripe distribution of the condensate magnons density across the stripe with the wave vector $2Q$\cite{Nowik-Boltyk 2012}. 

In these experiments the magnons were parametrically pumped by a microstrip resonator and analyzed by a Brillouin Light Scattering device. The frequency of the resonator was always smaller than $4\gamma H$, and the pumped magnons had a frequency less than the doubled gap $2\gamma H$. Therefore, the processes of decay for them were forbidden by energy conservation. The magnons' relaxation time $\tau_r$ was much shorter than their lifetime $\tau_l$ which is due to the dipolar interaction. Therefore, the magnons relaxed to a quasi-equilibrium state with non-zero chemical potential $\mu$. The magnitude of $\mu$ grew with the pumping power and reached the value of the energy gap $\Delta = 2\mu_B H$ at a critical magnitude of pumping. At pumping magnitude exceeding critical, the newly excited magnons went to the state with minimal energy and formed a condensate.Typical condensate densities were $10^{18}\sim 10^{19} cm^{-3}$. A typical relaxation time was $\tau_r ~ 100 \sim 200 ns$, the lifetime was $\tau_l ~ 3 \sim 10 \mu s$. 

All experiments were performed at room temperature that is smaller than the Curie temperature $T_c = 560$K. The appearance of the Bose-condensate at such a large temperature is possible because the pumped magnons do not leave the low-energy region, whereas the pumped energy flows to thermal magnons producing a negligibly small change in their temperature. 

\subsection*{Experiments}

In this section we briefly describe the crucial experiments by Borisenko \textit{et al.} and their results. 

\subsubsection*{Experiment I}
 In the first experiment \cite{Borisenko 2020-1} the sample of Yttrium Iron Garnet (YIG) was a thin square-shaped
film with the side $L=4mm$ and the thickness $d=5\mu m$
mounted on the interface of the dielectric resonator. The constant
magnetic field $H=600$ Oe parallel to one side of the film fixed
the spontaneous magnetization $M$ in the same direction. The experiment
was performed at room temperature. The magnitude of the spontaneous magnetization at this temperature
is $M=\frac{1375}{4\pi}=109.4G$.  A new feature of this experiment was a nanometer-thick golden strip of
the width $l=10\mu m$ surmounted under the sample perpendicularly
to the direction of the magnetic field $H$ and spontaneous magnetization
so that the central line of the strip divided the sample into two equal halves.
A permanent current
$I$ through the strip created a localized near the strip additional magnetic field $\mathbf{h}(z)$. 
Its $z$-component $h_{z}$ either antiparallel or parallel
to the field $H$ represents a potential well or a potential barrier
for magnons, respectively. The maximum magnitude of $h_{z}$ varied 
between 10 to 20 Oe. In all measurements the strong inequality
$\left|h_{z}\right|\ll H$ was satisfied. A schematic drawing of the
experimental device is shown in Fig. 1a.

\begin{figure}[!htb]
\centering
\includegraphics[width=8cm]{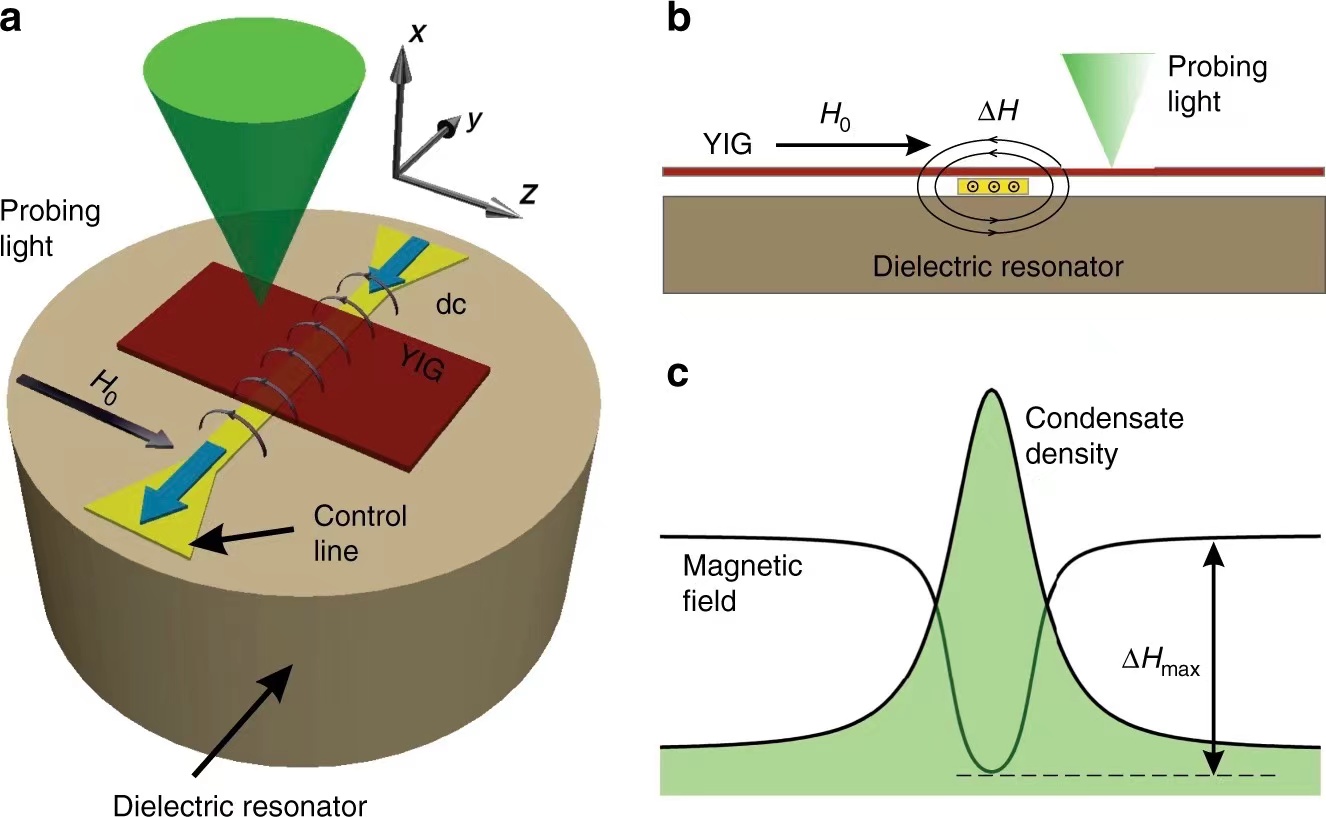}
\caption{Schematic picture of the experiment. Reprinted by permission from Macmillan Publishers Ltd: Nature Communications \cite{Borisenko 2020-1}, Copyright 2020. a) Schematic picture of the experimental device. The braun rectangle symbolizes the YIG film, the yellow strip is the golden strip carrying the electric current (blue arrow); the green cone symbolizes the Brilouin light scattering device; b) Schematic view of the magnetic field lines generated by the electric current in the golden strip; c) Schematic graphs of the condensate density and $z$-component of the magnetic field vs. coordinate $z$ along the direction of external magnetic field and spontaneous magnetization.}
 \label{fig:geometry}
\end{figure}

The main result obtained in the experiment I is the direct evidence
of the repulsion of the condensate magnons. To show that, the
experimenters focused on the case of the potential well for magnons. 
The density of condensate in this situation has a maximum at the
central line of the strip. The experimenters switched off the pumping 
and studied how the shape
of the density peak varies with time. The results are
shown in Fig. 2a, where profiles of the relative density $n\left(z,t\right)/n\left(0,t\right)$
are shown for several consequent moments of time starting from the
moment of switching off the pumping $t=0$. Fig. 2b shows the exponential
decay  of the condensate density with different decay times for different points of the initial
density profile. The Fig. 2a demonstrates that the width of the peak decreases
with time, i.e., with the average condensate density decreasing. This is direct evidence
of the magnons repulsion.
\begin{figure} 
\centering
  \includegraphics[width=0.75\textwidth]{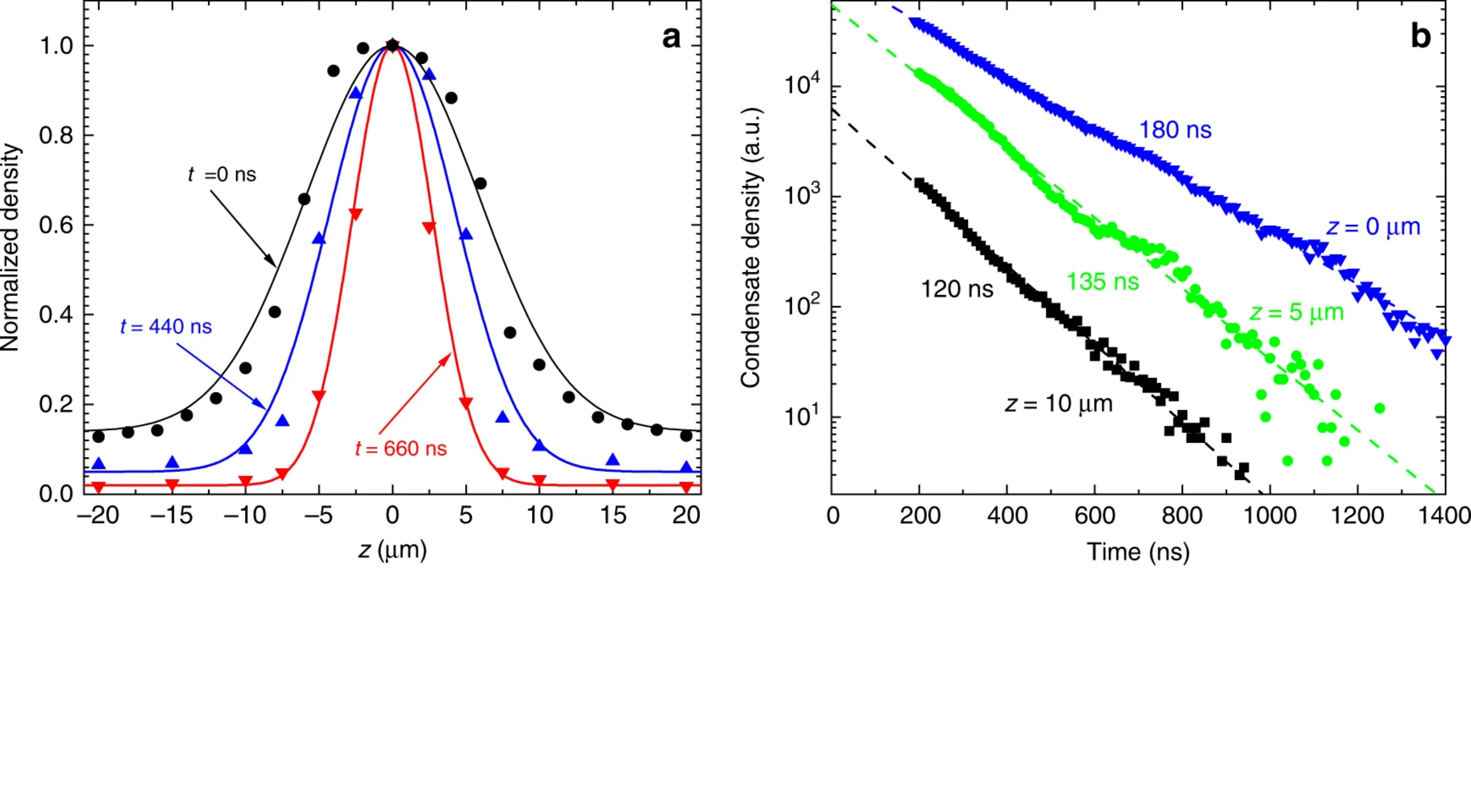}
  \caption{ (a) Normalized spatial profiles of the condensate density in a potential well with $\Delta H_{max}=-10$ Oe recorded at different delays after the microwave pumping is turned off at $t=0$ . Solid lines are guides for the eye. (b) Temporal dependence of the condensate density at different spatial positions. Dashed lines show the exponential fit of the experimental data with the corresponding effective decay times as indicated. Reprinted by permission from Macmillan Publishers Ltd: Nature Communications \cite{Borisenko 2020-1}, Copyright 2020.}
\label{fig:spectra2}
\end{figure}

\subsubsection*{Experiment II}
In the second experiment \cite{Borisenko 2020-2}, performed in the same geometry as the first one, a short pulse of the current through the golden strip created a pulse of magnetic field. The latter submitted
an additional amount of energy $\Delta\varepsilon$  to the condensate magnons. The imparted energy splits the
condensate initially distributed between the two minima into four moving
condensate clouds corresponding to the four points of the magnon spectrum
with energy $\varepsilon_{\min}+\Delta\varepsilon$ $\left(0<\Delta\varepsilon<\varepsilon_{\max}-\varepsilon_{\min}\right)$,
as shown in Fig. 3. These clouds have finite velocities. Let us enumerate the clouds in the order
they are located on the $z$-axis in the momentum space. Then their
velocities are  $-v_{1},v_{2},-v_{2},v_{1}$, where the velocities
are equal to derivatives $\frac{d\omega}{dk}$ in the four points
of the spectrum. Such a motion of the beams was observed in the experiment
II as it is shown in Fig. 3b,c. The clouds eventually smeared out since
the field in the pulse was non-uniform and $\Delta\omega$ depended
on $z$. The clouds displayed some asymmetry with respect to reflection
$k_{z}\Leftrightarrow-k_{z}$ that was not strong.
\begin{figure}[htb]
\centering
\includegraphics[width=0.6\textwidth]{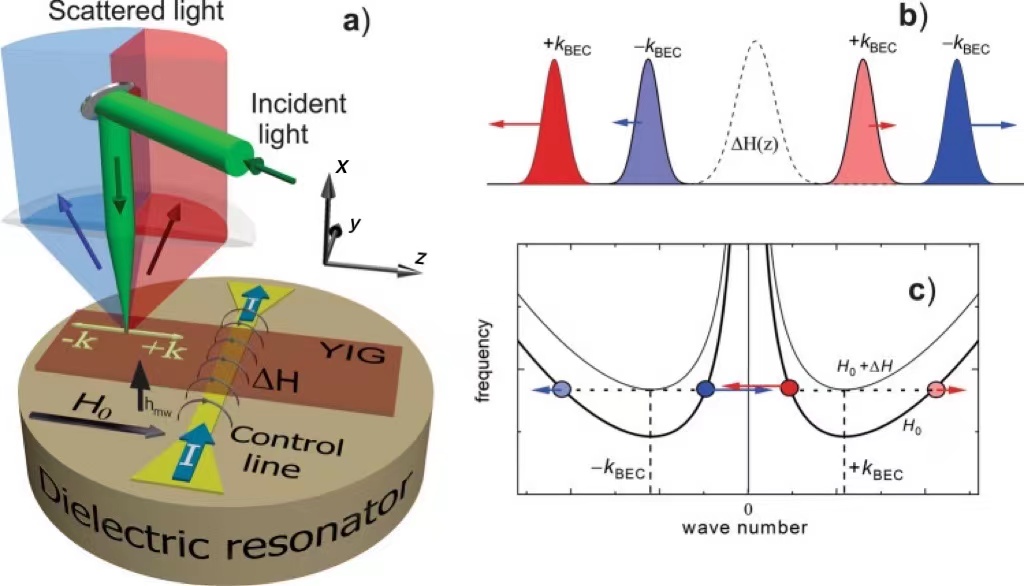}
\caption{Idea of the experiment II. (a) Schematic view of the experimental system. (b) After application of a pulse of localized magnetic field, BEC cloud is divided into four sub-clouds each of them moving
in direction, which is determined by the direction of the wavevector of the condensate and by the slope of the corresponding magnon spectral branch. (c) Magnon dispersion curves corresponding to the center of the metal strip, where magnetic field is equal to $H_0 + \Delta H$, and to a distant point, where the magnetic field is equal to $H_0$. Colored circles mark spectral states corresponding to the four split BEC clouds after they had moved away from center point. Arrows schematically indicate directions of the corresponding group velocities in the real space. Reprinted by permission from Macmillan Publishers Ltd:  Scientific Reports \cite{Borisenko 2020-2}, Copyright 2020.}
\end{figure}

\subsection*{Quasi-equilibrium theory. }

In the article \cite{Li 2013}
it was assumed that the weakly interacting gas of magnons relaxes
to the equilibrium state with non-zero chemical potential $\mu$.
It happens if the relaxation time $\tau_{r}$ is much less than
the lifetime of magnons $\tau_l$. In the basic approximation of non-interacting
magnons, the energy does not depend on the distribution of magnons
between the two minima of energy provided the total density of magnons
$n=n_{+}+n_{-}$ is fixed. This strong degeneracy is lifted by the magnons interaction.
The general form of the interaction free energy per unit volume is:
\begin{equation}
\begin{array}{c}
F_{int}=\frac{A}{2}\left(n_{+}^{2}+n_{-}^{2}\right)+Bn_{+}n_{-}+\\
C\sqrt{n_{+}n_{-}}\left(n_{+}+n_{-}\right)\cos\left(\phi_{+}+\phi_{-}\right)+\frac{D}{2}\left(n_{+}^{2}+n_{-}^{2}\right)\cos2\left(\phi_{+}+\phi_{-}\right)
\end{array}\label{eq:interaction}
\end{equation}
The calculation of coefficients $A$ and $B$ by Tupitsyn et al. \cite{Tupitsyn 2008} and
by F. Li et al. \cite{Li 2013} showed that for a thick film $d\gg\ell$, $A<0,$ $B>0$ and $\left|A\right|\ll B$
 The coefficients $C$ and $D$ found
first in the work by F. Li et al. and recalculated in \cite{Sun 2017} have magnitudes much smaller than $\left|A\right|$.
The coefficient $C$ and $D$ are nevertheless important. If they
are equal to zero, in the equilibrium state one of the two condensate
densities $n_{+}$ or $n_{-}$ is equal to zero. The coefficient $C$
makes both densities finite, though one of them is much less than
another. It also results in the phase trapping: the phase $\phi=\phi_{+}+\phi_{-}$
is equal to 0 or $\pi$ depending on the sign of the coefficient $C.$
A rather simple calculation of minimal free energy of condensate for
a thick film, at a fixed value of the total density $n$, gives the
two solutions for the ratios $\frac{n_{\pm}}{n}$. One of them is
\begin{equation}
\frac{n_{+}^{\left(1\right)}}{n}\approx1-\frac{C^{2}}{4B^{2}};\,\frac{n_{-}^{\left(1\right)}}{n}=\frac{C^{2}}{4B^{2}}\label{eq:asymmetric phase}
\end{equation}
The second solution differs from the first one by permutation of $n_{+}$
and $n_{-}$. These solutions are visibly strongly asymmetric. Thus,
in a thick film of YIG the interaction leads to a spontaneous violation
of symmetry between the two condensates: the density of one of them
becomes much larger than another. Their free energy is equal to $F_{int}\approx\frac{An^{2}}{2}<0$,
see also \cite{Tupitsyn 2008,Rezende 2009}. The attraction
makes homogeneous equilibrium state unstable. The collapse of the condensate 
is stopped by the growth of kinetic energy $K=\hbar^2(\nabla n)\hat{m}^{-1}(\nabla n)/(8n)$, where $\hat{m}$ is
the tensor of mass. It leads to formation of periodic distribution of the magnon density with the period $L\sim \hbar/\sqrt{8m\left| A\right| n} \sim15nm$. Still the total condensate energy remains negative.
This conclusion contradicts to the experiment I by Borisenko et al.
that has proved the repulsion of magnons. The existence of a periodic structure with so short period would quickly destroy the moving condensate clouds that was not observed in the experiment II. 

\subsection*{Results}

\subsection*{Slow inter-minima relaxation processes}
What was wrong in the quasi-equilibrium theory and why? The
only possible conclusion is that the complete equilibrium was not
established in the magnon system. The most probable reason is that
the inter-condensates relaxation time is much longer than the lifetime
of condensate magnons. 

Here we argue that it is indeed the case. The
processes that coherently flip a magnon from one minima to another are 
i) Compton scattering of the low-energy
magnon by a thermal magnon ii) the direct 4-th order processes
of coherent flip of condensate magnon interacting with a thermal magnon.
The Feynman diagrams of these two processes are shown in Figs. 4 and 5.
\begin{figure}[htb]
\centering
\includegraphics[width=0.5\textwidth]{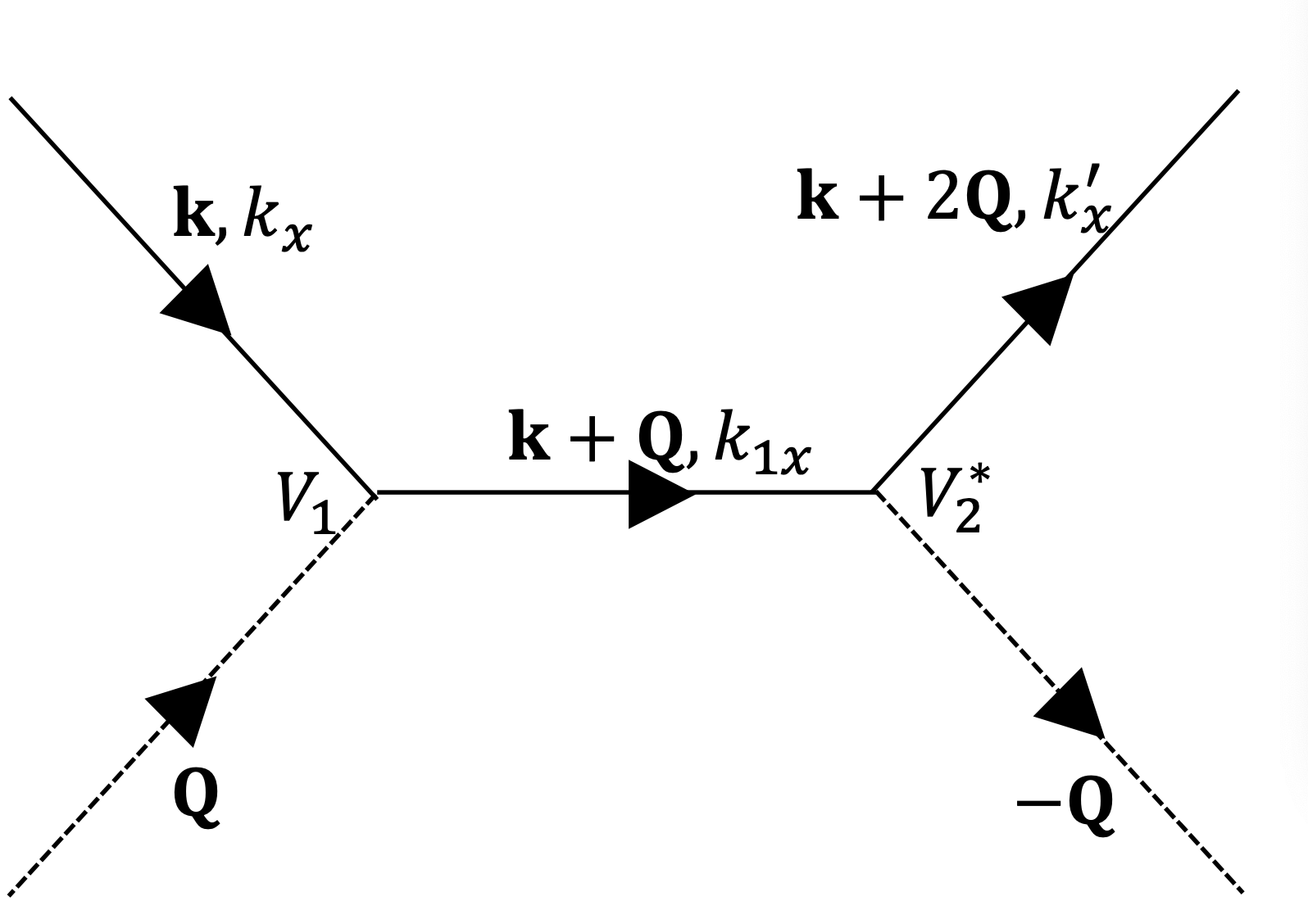}
\caption{Compton process for inter-minima relaxation.}
\end{figure}

In both processes the initial state contains the condensate magnon with in-plane momentum $\mathbf{Q}$ and perpendicular to plane momentum $\pi/d$ and a thermal magnon with much larger momentum components $\mathbf{k},k_x$. The condensate magnon inter-minimum transition processes turn them into a final pair of magnons $\mathbf{-Q}, \pi/d$ and $\mathbf{k}+2\mathbf{Q}, k^{\prime}_x$. The Compton process converts the initial state into an intermediate high-energy magnon with the in-plane momentum equal to the in-plane momentum of initial and final state (the intermediate magnon is off mass shell). In the 4-vertex process the same transition proceeds without intermediate magnon as the direct two-particle scattering. 

\subsubsection*{Rate of the Compton process}
The decay and merging processes are described by the vertex $V\left(\mathbf{k},k_x; \mathbf{k}^{\prime},k^{\prime}_x\right)$ that is completely due to the dipolar interaction. It can be represented as
\begin{equation}\label{dipolar vertex}
V(\mathbf{k},k_x;\mathbf{k}^{\prime},k^{\prime}_x)=\frac{8\pi^2(2\mu_B)^{3/2}M^{1/2}}{i\sqrt{Ad}}\frac{k_y k_z}{\mathbf{k}^2+k^{2}_{x}}\left[\frac{\cos \frac{(k_x-k^{\prime}_{x})d}{2}}{(k_x-k^{\prime}_{x})^2d^2-\pi^2}+\frac{\cos \frac{(k_x+k^{\prime}_{x})d}{2}}{(k_x+k^{\prime}_{x})^2d^2-\pi^2}\right],
\end{equation}
where $A$ is the area and $d$ is the thickness of the film. Employing this vertex, the amplitude of the Compton process is
\begin{equation}\label{compton-amplitude}
\mathcal{A}=d\int_{-\infty}^{\infty}\frac{dk_{1x}}{(2\pi)}\frac{V_1V_2^*\sqrt{N_{\mathbf{k+Q},k_{1x}}(N_{\mathbf{k+Q},k_{1x}}+1)}}{\varepsilon(\mathbf{k},k_x)+\varepsilon(\mathbf{Q})-\varepsilon(\mathbf{k+Q},k_{1x})+i\delta},
\end{equation} 
where $V_1=V(\mathbf{k},k_x,\mathbf{k}_1,k_{1x})$, $V_2=V\left(\mathbf{k}+2\mathbf{Q}, k^{\prime}_x,\mathbf{k}_1,k_{1x}\right)$
The inverse relaxation time or probability of the Compton process per unit time $1/\tau_C$ is given by:
\begin{equation}\label{relaxation time}
\frac{1}{\tau_C}=\frac{2\pi\left| N_{\mathbf{Q}}-N_{-\mathbf{Q}}\right|}{ \hbar}Ad^2\int \frac{d^3kd k^{\prime}_x}{(2\pi)^4}|\mathcal{A}|^2 N_{\mathbf{k},k_x}(N_{\mathbf{k},k_x}+1)\delta[\varepsilon(\mathbf{k},k_x)-\varepsilon(\mathbf{k+2Q},k_x')]
\end{equation}

Since $|\mathcal{A}|^2=(\Im \mathcal{A})^2+(\Re \mathcal{A})^2$, let us first estimate $\Im\mathcal{A}$:
\begin{equation}\label{ampl-imaginary}
\Im (\mathcal{A})=\frac{d}{2}\int_{-\infty}^{\infty} V_1V_2^{*}\sqrt{N_{\mathbf{k+Q},k_{1x}}(N_{\mathbf{k+Q},k_{1x}}+1)}
\delta\left( \varepsilon_{\mathbf{k+Q},k_{1x}} - \varepsilon_{\mathbf{k},k_{x}}\right) dk_{1x}.
\end{equation}
The energy of a thermal magnon is almost completely determined by exchange interaction. Therefore, it is propositional  to the square of its total momentum. Thus, $\Im\mathbf{A}$ is not zero only if the following constraint is satisfied:
\begin{equation}
\varepsilon(\mathbf{k},k_x)-\varepsilon(\mathbf{k+Q},k_{1x})=0 \rightarrow k_x^2-k_{1x}^2\approx 2k_z Q
\label{constraint-C}
\end{equation}
Then the square root of occupation numbers in eq. (\ref{ampl-imaginary}) can be estimated as 1 and the product $V_1V_2^*$ as
\begin{equation}
V_1V_2^*\approx  \frac{512\pi^4\mu_B^3 M}{81Ad}\frac{k_x^2k^{\prime 2}_x}{(k_zQd)^4}\cos \frac{(k_x-k_{1x})d}{2}\cos \frac{(k^{\prime}_{x}-k_{1x})d}{2}
\label{V-square-est}
\end{equation}

For the $\Re\mathcal{A}$ the constraint (\ref{constraint-C}) is invalid. However, the range of the variable $k_{1x}$ far from $k$ and $k^{\prime}$ so that $|k-k_{1x}|, |k^{\prime}-k_{1x}|\gg 1/d$, contributes negligibly small amount to the real part of integral (\ref{compton-amplitude}). Indeed, in this range of the variable $k_{1x}$ the vertexes $V_1,V_2$ decrease as $1/(kd)^2$ and strongly oscillate. Thus, in the region of $k_{1x}$ substantial for the integral (\ref{compton-amplitude}) the 
relations $|k_x - k_{1x-2k_zQ}|\sim 1/d\ll |2k_zQ|$ are satisfied. It means that $\Re\mathcal{A}$ has the same order of magnitude as $\Im \mathcal{A}$. 

We arrive at the estimate for the $\left|\mathcal{A}\right|^2$:
 \begin{equation}\label{ampl-im-est}
|\mathcal{A}|^2=\frac{32^2\pi^8 }{81^2A^2d^2}\frac{\mu_B^4}{k_T^2Q^8d^6\ell^4}|\cos \frac{(k_x-k^{\prime}_{x})d}{2}|^2.
\end{equation}
Then
\begin{equation}
\frac{1}{\tau_C}\approx \frac{2^7 \pi^8}{81^2}\frac{\mu_B^3}{k_T^3Q^8d^5\ell^6\hbar a^3M}\left(n_{\mathbf{Q}}-n_{-\mathbf{Q}}\right) \approx 3*10^{-7} s^{-1}.
\end{equation}
where $k_T=min\left(\sqrt{\frac{k_B T}{2\mu_B M}}\frac{1}{\ell},\frac{\pi}{a}\right)$
Then the Compton relaxation time is $\tau_c= 3.2*10^6 $s $\approx 38$ days.

\begin{figure}[htb]
\centering
\includegraphics[width=0.35\textwidth]{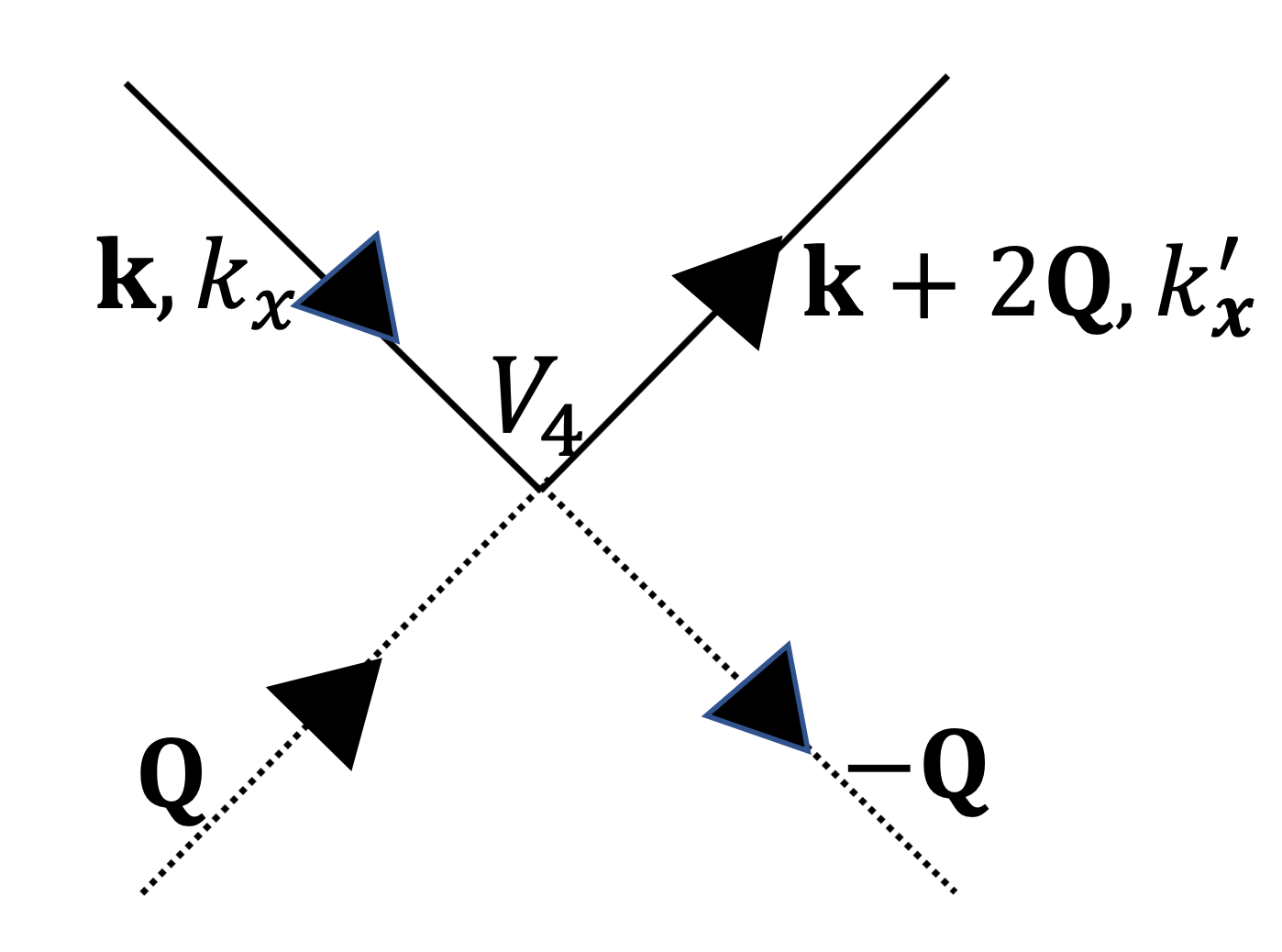}
\caption{4-vertex process.}
\end{figure}

\subsubsection*{Rate of the 4-vertex process}
For the rate of the 4-vertex process, we find:

\begin{equation}
\frac{1}{\tau_4}=\frac{2\pi}{\hbar}Ad^2\left(N_{\mathbf{Q}}-N_{-\mathbf{Q}}\right)\int\frac{d^3kdk^{\prime}_y}{(2\pi)^4}\vert\mathcal{A}_4\vert^2 N_{\mathbf{k},k_x}\left(1+N_{\mathbf{k},k_x}\right)\delta[\varepsilon(\mathbf{k},k_x)-\varepsilon(\mathbf{k+2Q},k_x^{\prime})]
\end{equation}
where the four-leg vertex $\mathcal{A}_4$ generated by both exchange and dipolar interaction is
\begin{equation}\label{A4}
\mathcal{A}_4=\frac{\mu_B^2\ell^2}{A}I; I=-\frac{8\pi^2k_x}{d^4}\left\{\frac{1}{k_zQ}+\frac{k_x^2Q^2}{(k_zQ)^3}\right\}\sin \frac{k_zQd}{k_x}
\end{equation}
The corresponding contribution to the probability of the condensate magnon inter-minima transition for $\vert n_{\mathbf{Q}}-n_{-\mathbf{Q}}\vert =10^{17} cm^{-3}$ is
\begin{equation}\label{tau-4}
\frac{1}{\tau_4}\approx \frac{32\pi^4\left(n_{\mathbf{Q}}-n_{-\mathbf{Q}}\right)\mu_B^3\ell^2}{\hbar k_T a^3 d^5 Q^2M}\approx 9.3*10^{-6} s^{-1}
\end{equation}
Thus, the 4-vertex process gives time of inter-minima relaxation $\tau_{i-m}=\tau_4 \approx 10^5$ s $\approx 1$ day. 
Both these processes have very small probability being proportional
to the fourth power of a small non-linearity.

Unfortunately, these estimates are not reliable. First of all, they depend on the difference of two condensate densities $n_{\mathbf{Q}} - n_{\mathbf{-Q}}$. We know reasonably well that their sum is a few 10$^{18}$ cm$^{-3}$. However, we do not know how small is the difference. In our estimates we used the value 5-10 times less than the sum. The second source of possible errors is high power of basic parameters entering the equations for $1/\tau_C$ and $1/\tau_4$. For the Compton process it is $\ell^6$, for the 4-vertex process it is $a^{-5}$. The estimates of $\ell$ by different authors differ by a value of the same order of magnitude. It means that the real value of inter-minima relaxation time may differ by the factor 32 from our estimate. The factor we denoted as $a^{-3}$ was an estimate of the magnitude of wave vectors significantly contributing the integral for $1/\tau_4$. It is also determined with the precision of factor 2.  

These uncertainties cannot, however, reduce the momentum-flip relaxation time to less than 1 hour that exceeds even the time of the experiment not speaking about the lifetime. It means that the equilibrium between the condensates in different minima is never established. Thus, the stationary state of the condensates is far from equilibrium. In this respect it is similar to the laser stationary state and, similarly to the laser,  the magnon condensate state may produce a coherent radiation of magnons \cite{Demokritov, Hillebrands}.

\subsection*{Conclusions: Anticipated properties of a stationary state with the condensate.}

Since the magnons repulse each other, the uniform stationary state of the
condensate does not collapse. We have shown that the inter-minima relaxation processes cannot establish equilibrium. However, it does not mean that the stationary state of the condensate is symmetric. The same processes in  a non-equilibrium state may be much stronger and sufficient for the dynamic spontaneous violation of the reflection symmetry. This problem needs the further theoretical study. %This prediction agrees with the inter-condensate
%interference structure of the condensate discovered by Novik-Boltyk
%et al. \cite{Nowik-Boltyk 2012}.

In real experiments it is difficult
to avoid a little asymmetry of the device that favors a slightly
asymmetric stationary state. Such a device asymmetry could explain
the asymmetry observed in the experiment II by Borisenko et al. If
the asymmetry is relatively small, then in eq. (\ref{eq:interaction})
the term $Bn_{+}n_{-}$ is dominant and positive, but unfortunately it completely disguises 
the possible dynamic spontaneous violation of the reflection symmetry.

Since the equilibrium for the inter-minima transitions is not established,
the complete theory of the stationary state with condensate requires a solution
of the Boltzmann kinetic equation for magnons together with the Gross-Pitaevskii equations
for the two condensates.  It can be done either by variational approach employing
the principle of maximal entropy production or by solution of a problem 
with proper initial conditions that asymptotically approaches a stationary
state. 

Such kinetic approach may resolve another discrepancy between the existing theories \cite{Bunkov-Volovik, Sun 2017} and experiment \cite{Demokritov, Hillebrands}.  The theory proofs that the pumped magnons are accumulated in the low-energy region assuming the temperature of accumulated magnons to be the same as initial temperature of the system (room temperature). Experimental data show that the low energy magnons have temperature about 3 times higher. The controversy may be resolved if the temperature is a slow-varying function of energy (momentum) that saturates to the room temperature of the system at some intermediate energy between $\mu_B H$ and the room temperature.

\subsection*{Acknowledgements}
We are thankful to Thomas Nattermann, Wayne Saslow, Fuxiang Li and Chen Sun who contributed to the theory on initial stages of this work. Our thanks are due to Sergei Demokritov, Vyacheslav Demidov, Igor Borisenko and Boris Divinsky whose experiments inspired this work.

\end{document}